# Mapping Unoccupied Electronic States of Freestanding Graphene by Angle-Resolved Low-Energy Electron Transmission

Flavio Wicki, Jean-Nicolas Longchamp, Tatiana Latychevskaia, Conrad Escher & Hans-Werner Fink

Physics Department of the University of Zurich, Winterthurerstrasse 190, CH- 8057 Zürich, Switzerland

**We report angle-resolved electron transmission measurements through freestanding graphene sheets in the energy range of 18 to 30 eV above the Fermi level. The measurements are carried out in a low-energy electron point source microscope, which allows simultaneously probing the transmission for a large angular range. The characteristics of low-energy electron transmission through graphene depend on its electronic structure above the vacuum level. The experimental technique described here allows mapping the unoccupied band structure of freestanding two-dimensional materials as a function of energy and probing angle, respectively in-plane momentum. Our experimental findings are consistent with theoretical predictions of a resonance in the band structure of graphene above the vacuum level** [V. U. Nazarov, E. E. Krasovskii, and V. M. Silkin, Physical Review B **87**, 041405 (2013)]**.**

## 1. Introduction

The electronic structure of a material is of great interest for a fundamental understanding of its mesoscopic properties and associated applications. While the occupied bands below the Fermi energy may be determined by different experimental methods like for example angle-resolved photoemission spectroscopy (ARPES) [1-4], the unoccupied bands are much more difficult to measure. However, experimental access to the unoccupied band structure above the vacuum level is possible to some extent with very low-energy electron diffraction (VLEED) spectroscopy [5-7], where the intensity of a reflected electron beam is measured as a function of electron energy and in-plane momentum. The electron reflectivity depends on the degree of coupling of the probing electron to empty states in the electronic structure of the sample. Another approach involves angle-resolved secondary electron emission (ARSEE), where excited electrons emitted from previously unoccupied states are identified in the fine structure of the secondary electron emission spectra [8-9].

Already in 1947, the linear dispersion of the electronic bands of graphene in the vicinity of the *K* point was predicted by Wallace using a tight binding model [10]. But it was only in 2004, when graphene became accessible to experimentalists [11] and since then, the energy states below and a few eV above vacuum level have been studied extensively both in theory and experimentally and are well understood by now [12-17]. However, only few studies are available concerning the higher lying electronic structure above the vacuum level. Ab-initio calculations show the existence of discrete states immersed in the continuous spectrum above the vacuum level [18-19]. Only recently, the existence of a special kind of scattering resonance in graphene originating from a strong coupling of the in-plane and perpendicular motions was predicted [18]. Ab-initio scattering theory [18,20] predicts that such resonances lead to a sharp transition from high to low transmission of an incident electron above respectively below the resonance energy. Measurements of the electronic structure above the vacuum level of single and multilayer graphene on SiC have been reported in a low-energy electron microscopy (LEEM) study based on the measuring principle of VLEED [21]. In an ARSEE experiment on graphene/Ni(111), spectral features were assigned to empty states in the graphene band structure by comparison with density functional theory calculations and a first experimental evidence of the aforementioned scattering resonance was provided [19]. To our knowledge, no experimental findings concerning the electronic structure of freestanding graphene above the vacuum level have been published so far.

2. Experimental Methods and Results

Here, we report the measurement of angle-resolved electron transmission through freestanding graphene using a low-energy electron point source (LEEPS) microscope [22]. A schematic of the experimental setup is shown in Fig. 1. A divergent beam of coherent low-energy electrons, field emitted from a sharp W(111) tip, is directed towards the graphene sample held at a distance of a few tens to hundreds of nanometers from the electron source. By means of a piezo positioner, the electron point source can accurately be positioned in front of the graphene sheet. The transmission signal through the graphene is recorded on a detector unit consisting of a micro-channel plate (MCP), a phosphor screen and a digital camera. The sensitive detector area has a diameter of 75 mm and is placed 47 mm behind the sample. Hence, transmission signal is captured within a full angle of 77°. Freestanding graphene over holes of 2 micrometer in diameter in a palladium coated silicon nitride membrane is prepared following the procedure described elsewhere [23].

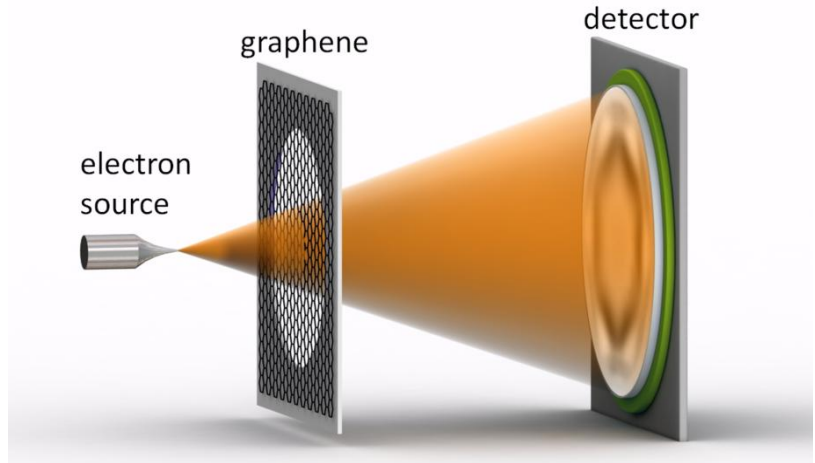

FIG. 1. Schematics of the LEEPS microscope setup. A sharp W(111) tip acts as a field emitter source of a divergent electron beam. The electron wave is scattered off the sample placed in close proximity to the source. The result of this scattering experiment is captured at a distant detector.

Single layer graphene has proven to be a suitable sample substrate with a transparency of the order of 70% for electron energies of 50 to 250 eV typically employed in LEEPS holography experiments [24-25]. However, when lowering the electron energies to about 30 eV and below, we find a rather strong angular dependence of the electron transmission through freestanding graphene. The transmission images from a single layer graphene sheet show 6-fold symmetric patterns. Moreover, the orientation of the patterns depends on the one of the crystal lattice of the particular graphene domain. This observation is exemplified in Fig. 2, where LEEPS microscopy images are shown recorded with 22 eV electrons while moving the field emission tip across the boundary of two adjacent domains in a single layer graphene sheet. Apparently, the orientation of the 6-fold symmetric pattern in the transmission image changes according to the change of the domain orientation, as indicated in Fig. 2(d) and 2(f). The transmission image shown in Fig. 2(b) is recorded with the tip placed right in front of the domain boundary. Elastic scattering of clusters decorating the domain boundary leads to the formation of a hologram. Note that part of the 6-fold symmetric pattern associated with the transmission through graphene is apparent with different orientation on either side of the domain boundary. At first sight, one might be tempted to consider the appearance of the hexagon being due to some sort of diffraction pattern. However, the condition of a plane or slightly divergent wave creating a far-field diffraction pattern are not given and moreover, such pattern, probing two graphene domains at once would resemble a superposition of two similar 6-fold symmetric patterns. The diameter of the probed area varies with the source to sample distance and amounts to about hundred nanometers in Fig. 2.

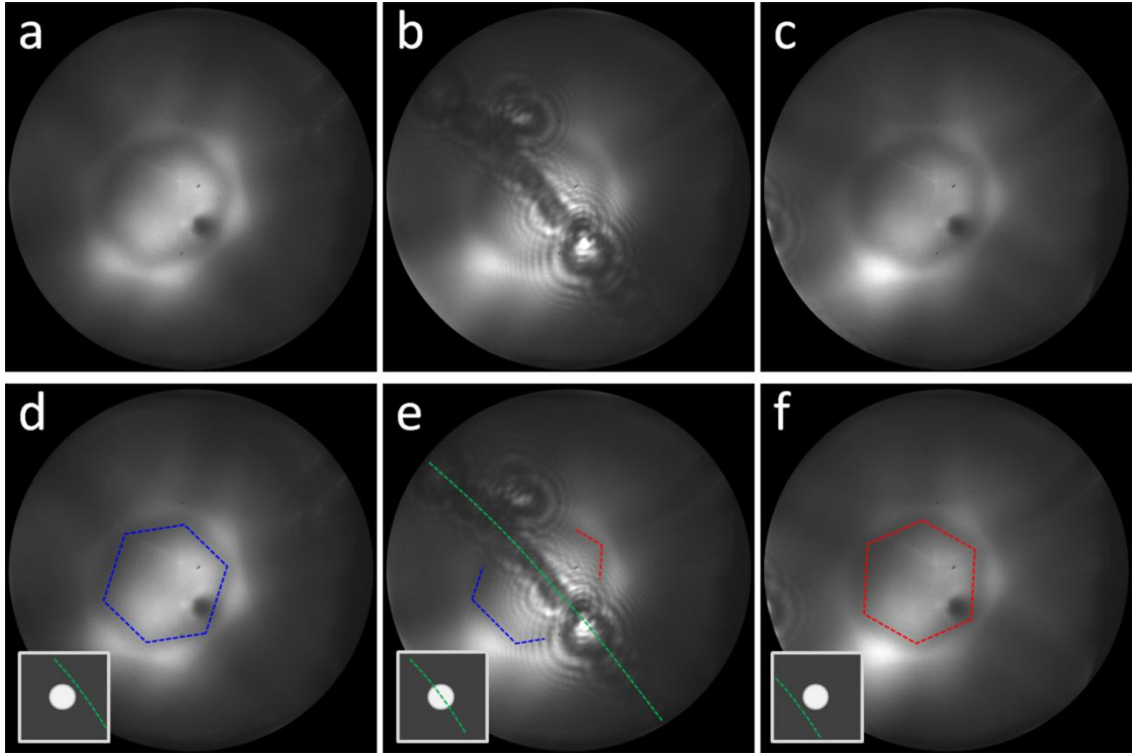

FIG. 2. Low-energy electron transmission images of single layer freestanding graphene recorded while moving the field emission tip across a domain boundary. (a) Transmission image recorded with the tip on the left side of the boundary. (b) Image recorded with the tip in front of the boundary. Clusters decorating the boundary lead to the formation of an in-line hologram. (c) Transmission image recorded on the right side of the domain boundary. (d)-(f) The same images as in (a)-(c) but with blue and red coloured dashed lines to guide the readers eye to the different orientations of the 6-fold symmetric pattern, which depend on the orientation of the crystal lattice of the particular graphene domain. In (e) the green dashed line indicates the domain boundary. The bright discs in the sketches at the bottom left corner of (d)-(f) indicate the position of the probed sample area with respect to the domain boundary. The diameter of the probed sample area amounts to about 100 nm.

In the transmission images presented here, the sample is probed at once under a large angular range and at constant kinetic energy of the incident electrons. The in-plane momentum component $k_{//}$ of a probing electron is determined by the angle $\theta$ as illustrated in Fig. 3(a). A single transmission measurement thus corresponds to a constant energy map in $T(\mathbf{k}_{//},E)$, the electron transmission as a function of in-plane momentum and energy. Since the transmission probability for a low-energy electron depends on its specific coupling to the electronic states of the sample, such a measurement allows drawing conclusions about the electronic structure $E(\mathbf{k})$ of the sample.

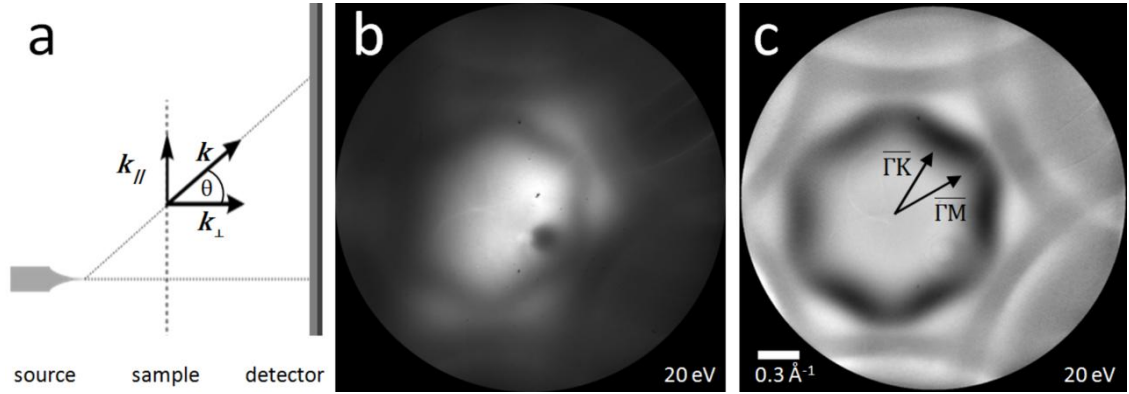

FIG. 3. Principle of in-plane momentum resolved transmission measurements in the LEEPS microscope. (a) Schematics illustrating the determination of the in-plane momentum components of the incident electrons. Electrons are field-emitted from a sharp tip and arrive at the freestanding graphene sheet under a large angular range. The in-plane momentum of an electron in the graphene plane is related to the probing angle $\theta$ and the total momentum $\hbar k = \sqrt{2 m_e E_{kin}}$. While the kinetic energy is determined by the tip bias potential and the work function of the sample, the angle can be determined from the position where the electron is recorded in the detector plane. (b) Raw data image of the transmission intensity recorded with an electron energy of 20 eV. The transmission intensity shows a 6-fold symmetry in accordance with the reciprocal crystal lattice of graphene, the background intensity varies due to anisotropic emission from the field emitter tip. The small dark spots are due to detector defects, the larger dark spot in the bottom-right corner of the dark hexagonal shape is the blind spot of the MCP where electrons enter nearly parallel to the channels and are less amplified. (c) Background corrected transmission intensity of (b)

interpolated to $k_{//}$-space. The arrows indicate the directions along the high symmetry axes $\overline{\varGamma M}$ and $\overline{\varGamma K}$ in the Brillouin zone.

The energy $E$ of the probing electrons with respect to the Fermi level of the graphene sample is given by the tip potential $E=eU_0$, and their kinetic energy $E_{kin}$ behind the sample is given by $E_{kin}=eU_0-\varPhi_s$, with $\varPhi_s$ being the work function of the sample. The width of the energy distribution of the emitted electrons amounts to roughly 250 meV, typical for field emission from a (111)-oriented tungsten emitter [26-27]. The in-plane wave vector magnitude of the electrons is conserved in a sample/vacuum transition and is given by $k_\parallel = k \cdot \sin\theta$ with the total momentum $\hbar k = \sqrt{2m_e E_{kin}}$ and the probing angle $\theta$.

A raw data transmission image recorded with an electron energy of 20 eV is shown in Fig. 3(b). We correct for the anisotropic emission of the electron source by dividing the transmission images with an approximate background image calculated as the average of a series of transmission images recorded with different energies (see Fig. S1 (c) in Supplemental Material [28]). In Fig. 3(c) the background corrected data of Fig. 3(b) are shown, interpolated to $k_{//}$-space [29]. The arrows indicate the directions along $\overline{\varGamma M}$ and $\overline{\varGamma K}$ of the Brillouin zone. The orientation of the Brillouin zone is determined by a comparison with the first order diffraction disks apparent when larger electron energies are used (see Fig. S1 (a) and (b) in Supplemental Material [28]).

A relatively high transmission is observed for low $k_{//}$-values. For increasing $k_{//}$-values a dark hexagonal region of low transmission oriented with the edges pointing in the $\overline{\varGamma M}$ direction passes into a region of high transmission with a larger extent in the $\overline{\varGamma K}$ direction. The positions and widths are in agreement with the lowest branch of the scattering resonance predicted by Nazarov et al. (see Fig. 4 in Ref. [18]). We therefore conclude that this is the first experimental observation of this scattering resonance in freestanding graphene.

The transitional region from low to high transmission is surrounded by a region of reduced transmission in the form of a six-pointed star with nearly constant width in $k_{//}$-space. This again is in good agreement with ab-initio calculations [18][30].

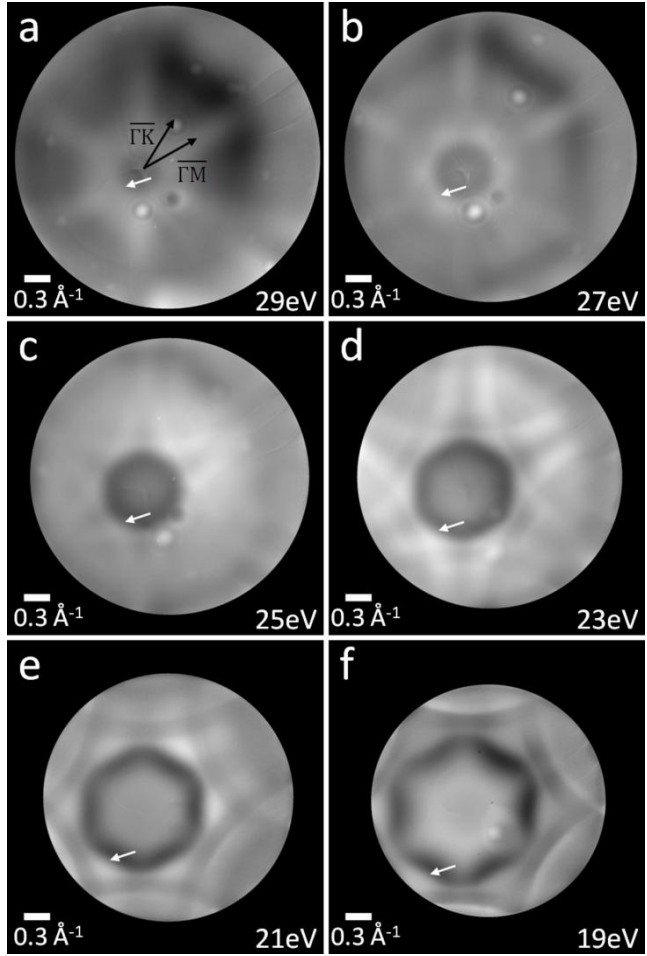

FIG. 4. Background corrected transmission intensity in linear grayscale as a function of $k_{//}$ recorded with electron energies ranging from 29eV down to 19eV. The orientation of the Brillouin zone is indicated with according black arrows in (a). Each image corresponds to a constant energy map of the scattering band structure $T(k_{//},E)$. The few bright spots are due to impurities on the graphene (e.g. visible in (a)). The transmission probability for a probing electron depends on its specific coupling to the unoccupied states of the sample which leads to signatures in the constant energy maps of $T(k_{//},E)$ with a 6-fold symmetry. The pronounced transition from low to high transmission (marked with white arrows) is associated with a scattering resonance in compliance with ab-initio predictions [18].

In Fig. 4, background corrected transmission images are shown for electron energies of 29 eV down to 19 eV in steps of 2 eV. The probed sample area, which is determined by the tip to sample distance, is about 300 nm in diameter for the measurement recorded with 29 eV shown in Fig. 4(a). When lowering the electron energy respectively the bias potential, the tip to sample distance is reduced to maintain a constant emission current of a few nanoamperes. Accordingly, the probed sample area gradually shrinks to about 50 nm in diameter at 19 eV as shown in Fig. 4(f). Although the background correction provides transmission intensities in arbitrary units only, the intensities can reasonably be compared among different energies.

The transition from low to high transmission associated with the scattering resonance can clearly be distinguished for all energies. With decreasing energy this transition shifts to higher $k_{//}$-values and the dispersion is in good agreement with theoretical predictions [18].

The dispersion of the low transmission branch of the scattering resonance can easily be recognized in Fig. 5, where the transmission intensity along the high symmetry directions $\overline{\varGamma M}$ and $\overline{\varGamma K}$ is shown as a function of $E$ and $k_{//}$. The intensity is extracted along the direction of the arrows indicated in Fig. 4(a) from a measurement series recorded with electron energies from 30 eV down to 18 eV in 1 eV steps (see Movie 1 in Supplemental Material [28]). The transmission intensity along the high symmetry lines is in qualitative agreement with measurements of electron reflection from graphene on SiC [21]. A reduced transmission is measured for regions in ($E,k_{//}$)-space where a relatively high reflection has been reported.

In the transmission intensity distributions at electron energies of 29 eV and 27 eV shown in Figs. 4(a) and 4(b), the transmission at $k_{//}$-values above the scattering resonance is larger along the $\overline{\varGamma M}$ direction than along the $\overline{\varGamma K}$ direction, in accordance with theoretical predictions. This gradually changes when the energy is lowered and for 23 eV, shown in Fig. 4(d), the transmission along the $\overline{\varGamma M}$ direction is reduced. In the transmission images recorded at the very low electron energies of 21 eV and 19 eV, there is an additional region of lower transmission appearing for $k_{//}$-values above the scattering resonance. This outer domain of low transmission can be clearly recognized as a six-pointed star enclosing the hexagonal pattern in Figs. 4(e) and 4(f).

In general, the characteristics observed here are in excellent agreement with the ab-initio predictions regarding their position in energy and in-plane momentum space. However, the relative transmission intensities deviate from the calculations. For example, the highest transmission probability is predicted to be at $k_{//}$-values above the scattering resonance for all energies. In contrary to that, the experimental findings suggest a larger transmission in the centre of the Brillouin zone at an energy of 19 eV as shown in Fig. 4(f). We attribute this discrepancy to inelastic processes giving rise to enhanced absorption at higher $k_{//}$-values. Such effects have not been taken into account for the ab-initio calculations [18].

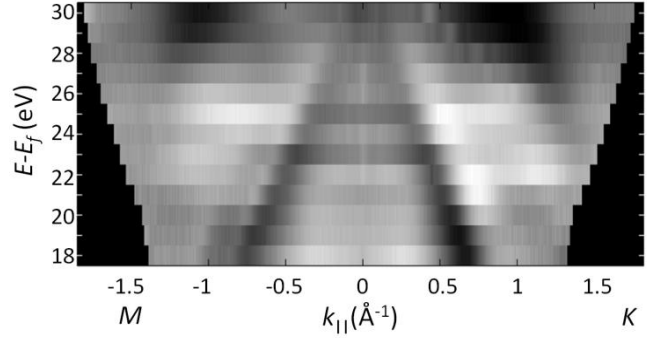

Fig 5. Plot of the background corrected transmission intensity as a function of the electron energy and the $k_{//}$-values in $\overline{\varGamma M}$ and $\overline{\varGamma K}$ direction of the Brillouin zone. The intensity was extracted in the direction of the arrows indicated in Fig. 4(a) for transmission images recorded with energies from 30 eV down to 18 eV in steps of 1 eV. The dispersion of the low transmission signature associated with the scattering resonance is in good agreement with theoretical predictions [18].

3. Conclusions

To summarize, we carried out angle-resolved electron transmission measurements on freestanding graphene revealing scattering resonances predicted by theory. Using the setup of a LEEPS microscope, it is possible to record constant energy maps of the transmission function $T(k_{//},E)$ in a fraction of a second. Unoccupied states can thus be mapped in all directions of the Brillouin zone at once, rather than having to scan along a certain direction of the Brillouin zone and measuring point for point as in common VLEED or ARSEE setups. Moreover, the probing area can be as small as a few tens of nanometers in diameter and thus allows probing of freestanding two-dimensional materials, which can often be prepared on the micrometer scale only.

The technique described here can also be applied to study other two-dimensional materials provided that they are sufficiently transparent to low-energy electrons and can be prepared freestanding. Possible candidates for further studies are few layer graphene sheets, single layer hexagonal boron nitride/graphene compounds, exfoliated single layer molybdenum disulphide [31] or heterostructures of graphene and transitional metal dichalcogenides [32].

**Acknowledgment:**

The work presented here has been financially supported by the Swiss National Science Foundation (SNF).

**Supplemental Material:**

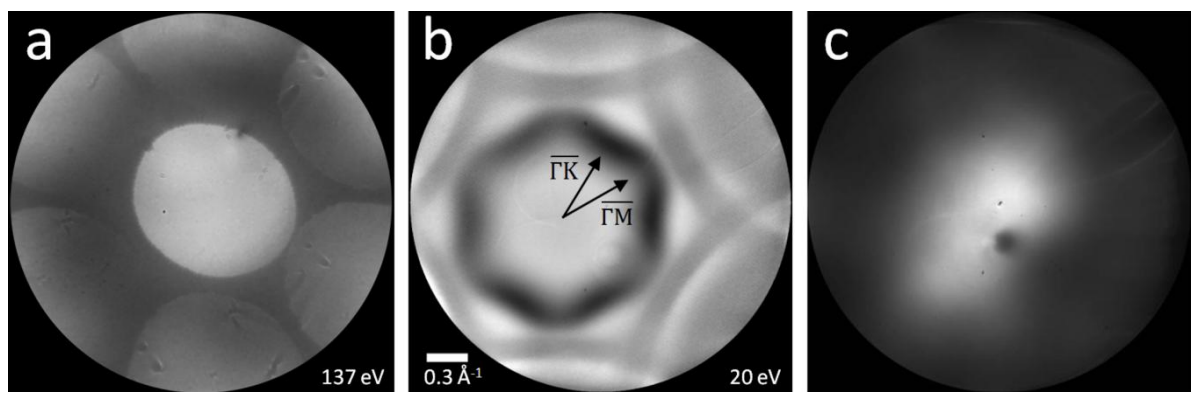

FIG. S1. (a) First order diffraction disk image used to determine the orientation of the Brillouin zone. (b) Background corrected transmission image at 20 eV with arrows to indicate the orientation of the Brillouin zone. (c) Background intensity calculated as the average of a series of transmission images recorded at different energies.